\documentclass[12pt]{article}
 
\usepackage{graphicx}
\usepackage{amsfonts}
\usepackage{amsthm}
\usepackage{amsmath}
\usepackage{bbm}
\usepackage[top=2.75cm, bottom=2.75cm, left=2.75cm, right=2.75cm]{geometry} 

\newcommand{\Section}[1]{\section{#1} \setcounter{equation}{0}} 
\newcommand{\beq}{\begin{equation}} 
\newcommand{\eeq}[1]{\label{#1}\end{equation}}
\newcommand{\ber}{\begin{eqnarray}} 
\newcommand{\eer}[1]{\label{#1}\end{eqnarray}}

\newcommand{\ub}{\bar{u}}
\newcommand{\zb}{\bar{z}}
\newcommand{\xb}{\bar{x}}
\newcommand{\yb}{\bar{y}}
\newcommand{\zetab}{\bar{\zeta}}
\newcommand{\Ut}{\tilde{\Upsilon}}
\newcommand{\Uts}{\bar{\tilde{\Upsilon}}}
\newcommand{\Us}{\bar{\Upsilon}}
\newcommand{\U}{\Upsilon}
\newcommand{\Ub}{\bar{\Upsilon}}
\newcommand{\Utb}{\bar{\tilde{\Upsilon}}}

\begin{document}
\renewcommand{\theequation}{\thesection.\arabic{equation}}  
\setcounter{page}{0}

\thispagestyle{empty}

\bigskip
\begin{center} \LARGE
{\bf A twistor sphere of generalized K\"ahler potentials on hyperk\"ahler manifolds}
 \\[12mm] \normalsize
{\bf Malte~Dyckmanns\footnote{E-mail: malte.dyckmanns@udo.edu}} \\[8mm]
{\small\it ~\\
C.N.Yang Institute for Theoretical Physics, Stony Brook University, \\
Stony Brook, NY 11794-3840,USA\\
~\\~\\}
\end{center}
\vspace{10mm}
\centerline{\bfseries Abstract} 
\bigskip

\noindent  
We consider the generalized K\"ahler structures $(g,J_+,J_-)$ that arise on a hyperk\"ahler mani-fold $(M,g,I,J,K)$ when we choose $J_+$ and $J_-$ from the twistor space of $M$. We find a relation between semichiral and arctic superfields which can be used to determine the generalized K\"ahler potential for hyperk\"ahler manifolds whose description in projective superspace is fully understood. We use this relation to determine an $S^2$-family of generalized K\"ahler potentials for Euclidean space and for the Eguchi-Hanson geometry. Cotangent bundles of Hermitian symmetric spaces constitute a class of hyperk\"ahler manifolds where our method can be applied immediately since the necessary results from projective superspace are already available. As a non-trivial higher-dimensional example, we determine the generalized potential for $T^\ast\mathbb{C}P^n$, which generalizes the Eguchi-Hanson result.
\bigskip
\bigskip ~
\eject
\normalsize
\eject

\section{Introduction}
Hyperk\"ahler manifolds admit various generalized K\"ahler structures. The corresponding generalized K\"ahler potentials can be used to reconstruct the hyperk\"ahler geometry. These generalized potentials are in general quite different from the ordinary K\"ahler potential and thus provide a new way of studying hyperk\"ahler geometry and finding hyperk\"ahler metrics. To gain insight into this new way of looking at hyperk\"ahler geometry, one first needs to study some examples. In this paper, we make use of the twistor space of hyperk\"ahler manifolds to develop a general method for determining their generalized K\"ahler potentials and we also explicitly work out some examples.

In the next section, we review the relevant features of generalized K\"ahler geometry in its bihermitian formulation. This geometry involves two complex structures $J_+$, $J_-$ on a Riemannian manifold $(M,g)$ and can locally be described by a generalized K\"ahler potential. In this paper, we consider the case where the kernel of $[J_+,J_-]$ is trivial. Then the potential is defined as the generating function for a symplectomorphism between coordinates $(x_L,y_L)$ and $(x_R,y_R)$ that are holomorphic w.r.t. $J_+$ and $J_-$ respectively. Generalized K\"ahler geometry was initially found as the target space geometry of $2D$ $\mathcal{N}=(2,2)$ supersymmetric sigma models, where the potential is the superspace Lagrangian and the coordinates $x_L,x_R$ describe semichiral superfields.

In section \ref{sectionhyper}, we review aspects of hyperk\"ahler geometry and its twistor space. We parametrize the twistor sphere of complex structures by a complex coordinate $\zeta$ and introduce holomorphic Darboux coordinates $\U(\zeta),\Ut(\zeta)$ for a certain holomorphic symplectic form. This construction is relevant for the projective superspace description of $2D$ $\mathcal{N}=(4,4)$ sigma models, where $\U,\Ut$ are arctic superfields.

Using its twistor space, a hyperk\"ahler manifold can be seen as a generalized K\"ahler manifold in various ways while keeping the metric fixed. In section \ref{sectionstructures}, we consider a two-sphere of generalized K\"ahler structures on a hyperk\"ahler manifold and express the  coordinates $x_L,x_R,y_L,y_R$ in terms of $\U,\Ut$. This enables us to determine the generalized K\"ahler potential on a hyperk\"ahler manifold if we can find the decomposition of the arctic superfields $\U,\Ut$ in terms of their $\mathcal{N}=(2,2)$ components, i.e. in terms of coordinates on $M$.

In section \ref{sectionfour}, we consider four-dimensional hyperk\"ahler manifolds and explicitly determine the partial differential equations that the coordinates describing those arctic superfields have to fulfill. In section \ref{sectioneuclid}, we determine the potential for Euclidean space, where the differential equations for $\U,\Ut$ are easy to solve. In section \ref{sectioneguchi}, we look at the Eguchi-Hanson metric, where the relevant coordinates $\U,\Ut$ have been found previously in \cite{properties}. We give an explicit expression for the $S^2$-family of generalized K\"ahler potentials for this geometry, which belongs to the family of gravitational instantons and is thus of interest to physicists.

The Eguchi-Hanson geometry lives on the cotangent bundle of $\mathbb{C}P^1$. For hyperk\"ahler structures on cotangent bundles over arbitrary K\"ahler manifolds, projective superspace can be used to determine the coordinates $\U,\Ut$. This has been done in particular for all Hermitian symmetric spaces. In section \ref{sectionproj}, we review this procedure and as a non-trivial higher-dimensional example, we use the results for $T^\ast\mathbb{C}P^n$ to determine its generalized K\"ahler potential, which generalizes the Eguchi-Hanson result.

In an appendix, we extend our results from section \ref{sectionstructures} and consider the full $S^2\times S^2$-family of generalized K\"ahler structures on a hyperk\"ahler manifold, i.e. we let both $J_+$ and $J_-$ be an arbitrary point on the twistor sphere of complex structures. We also give the explicit potential depending on two complex parameters $\zeta_+$,$\zeta_-$ for the simplest hyperk\"ahler manifold, namely for Euclidean space.

\section{The generalized K\"ahler potential}
Generalized K\"ahler geometry first appeared in the study of 2D $\mathcal{N}=(2,2)$ nonlinear $\sigma$-models \cite{gates} and was later rediscovered by mathematicians as a special case of generalized complex geometry \cite{gualtieri}. In its bihermitian formulation, generalized K\"ahler geometry consists of two (integrable) complex structures $J_+$, $J_-$ on a Riemannian manifold $(M,g)$, where the metric is hermitian with respect to $J_+$ and $J_-$. Furthermore, the forms $\omega_\pm:=gJ_\pm$ have to fulfill \cite{linearizing}
\beq 
d^c_+\omega_++d^c_-\omega_-=0,\quad dd^c_+\omega_+=0,
\eeq{integrability}
where $d^c_\pm=i(\bar{\partial}_\pm-\partial_\pm)$. This allows us to define the closed three-form $H:=d^c_+\omega_+=-d^c_-\omega_-$, whose local two-form potential we denote by $B$ ($H=dB$).
In general, $\omega_\pm$ is not closed and thus $(M,g,J_\pm)$ is not K\"ahler. In this paper, we will however consider the case where $H=0$. Then $\partial_\pm\omega_\pm$, $\bar{\partial}_\pm\omega_\pm$ have to vanish separately, so $d\omega_\pm=0$, i.e. $(M,g,J_\pm)$ is K\"ahler.

In \cite{offshell} it was shown that like ordinary K\"ahler geometry, generalized K\"ahler geometry is locally described by a single function, the generalized K\"ahler potential. On a generalized K\"ahler manifold, one can define the Poisson structure $\sigma:=[J_+,J_-]g^{-1}$ \cite{hitchinP}. Here, we consider the case where $[J_+,J_-]$ is invertible and recall how the generalized K\"ahler potential is defined in this case \cite{potential}:

Inverting $\sigma$ gives
\beq \Omega_G:=\sigma^{-1}=g[J_+,J_-]^{-1}, \eeq{omegag}
which is a real, closed and non-degenerate two-form that fulfills $J_\pm^T\Omega_GJ_\pm=-\Omega_G$ \cite{hitchinP}, i.e. it is a real holomorphic symplectic form both w.r.t. $J_+$ and w.r.t. $J_-$. This means that $\Omega_G$ can be split into the sum of a $(2,0)$- and a $(0,2)$-form both w.r.t. $J_+$ and w.r.t. $J_-$:
\beq \Omega_G=\Omega_+^{(2,0)}+\Omega_+^{(0,2)}=\Omega_-^{(2,0)}+\Omega_-^{(0,2)}, \eeq{plusminus}
where $\bar{\partial}_\pm\Omega_\pm^{(2,0)}=0$ and $\Omega_\pm^{(0,2)}=\overline{\Omega_\pm^{(2,0)}}$ (here the complex conjugate is taken w.r.t. $J_+$ and $J_-$ respectively).

One then introduces Darboux coordinates $x^p_L$ and ${y_L}_p$, holomorphic w.r.t. $J_+$, for $\Omega_+^{(2,0)}$; and $x^p_R$ and ${y_R}_p$, holomorphic w.r.t. $J_-$, for $\Omega_-^{(2,0)}$ ($p=1,...,n$, where $dim_\mathbb{R}M=4n$) \cite{potential}. Then  
\ber \Omega_G&=&\Omega_+^{(2,0)}+\Omega_+^{(0,2)}=dx_L^p\wedge {dy_L}_p+d\xb_L^p\wedge {d\yb_L}_p, \nonumber \\
 \Omega_G&=&\Omega_-^{(2,0)}+\Omega_-^{(0,2)}=dx_R^p\wedge {dy_R}_p+d\xb_R^p\wedge {d\yb_R}_p, \eer{asdf}
i.e. the coordinate transformation from $\{x_L,\xb_L,y_L,\yb_L\}$ to $\{x_R,\xb_R,y_R,\yb_R\}$ is a symplectomorphism (canonical transformation) preserving $\Omega_G$. It is thus described by a generating function $P(x_L,x_R,\xb_L,\xb_R)$ such that (omitting indices from now on)
\beq \frac{\partial P}{\partial x_L}=y_L,\quad \frac{\partial P}{\partial x_R}=-y_R,\quad \frac{\partial P}{\partial \xb_L}=\yb_L,\quad \frac{\partial P}{\partial \xb_R}=-\yb_R. \eeq{lrRelation}

This generating function is the generalized K\"ahler potential\footnote{We can choose to let the potential depend on other combinations of old and new coordinates as well. The potentials corresponding to the four different choices of variables are then related via Legendre transforms. In previous papers, the roles of $x_R$ and $y_R$ were interchanged. However, formulas in previous papers for reconstructing $g$, $J_+$, $J_-$ and $B$ from the potential remain unchanged when using our convention.} and can be used to locally reconstruct all the geometric data of generalized K\"ahler geometry \cite{potential}, i.e. the two complex structures $J_+$, $J_-$, the metric $g$ and the $B$-field. It also turns out to be the superspace Lagrangian for the $\mathcal{N}=(2,2)$ $\sigma$-models that led to the discovery of generalized K\"ahler geometry \cite{offshell}.

\section{Hyperk\"ahler manifolds and their twistor spaces}\label{sectionhyper}
In this paper, we consider generalized K\"ahler structures $(g,J_+,J_-)$ on a hyperk\"ahler mani-fold $M$ and investigate their generalized K\"ahler potentials. For the choice of the two complex structures $J_+$ and $J_-$, we will make use of the twistor space $\mathcal{Z}=M\times S^2$ of $M$.

Hyperk\"ahler manifolds appear for instance as the target spaces for hypermultiplet scalars in four-dimensional nonlinear $\sigma$-Models with rigid $\mathcal{N}=2$ supersymmetry on the base space \cite{freedman}. In geometric terms, they are described by the data $(M,g,I,J,K)$, where $g$ is a Riemannian metric on $M$ that is K\"ahler with respect to the three complex structures $I,J,K$, which fulfill the quaternion algebra (i.e. $IJ=K=-JI$). In fact, there exists a whole two-sphere of complex structures on $M$ with respect to which $g$ is a K\"ahler metric, namely $(M,g,\mathcal{J}=v_1I+v_2J+v_3K)$ is K\"ahler for each $(v_1,v_2,v_3)\in S^2$. Using (the inverse of) the stereographic projection, we parametrize this family of complex structures on $M$ in a chart of $S^2$ including the north-pole by a complex coordinate $\zeta$:
\beq \mathcal{J}(\zeta):=v_1(\zeta)I+v_2(\zeta)J+v_3(\zeta)K:=\frac{1}{1+\zeta\zetab}\left[(1-\zeta\zetab)I+(\zeta+\zetab)J+i\,(\zetab-\zeta)K\right]. \eeq{J}

We define the complex two-forms
\beq \omega^{(2,0)}:=\omega_2+i\omega_3,\quad\quad\omega^{(0,2)}:=\omega_2-i\omega_3; \eeq{omegapm}
where $\omega_1=gI$, $\omega_2=gJ$, $\omega_3=gK$ are the three K\"ahler forms. Then for each $\zeta\in \mathbb{C}$
\beq \Omega_H(\zeta):=\omega^{(2,0)}-2\zeta\omega_1-\zeta^2\omega^{(0,2)} \eeq{OmegaH}
turns out to be a holomorphic symplectic form with respect to the complex structure $\mathcal{J}(\zeta)$ \cite{hitchin}. In particular, $\omega^{(2,0)}=\Omega_H(\zeta=0)$ is a $(2,0)$-form w.r.t. $I=\mathcal{J}(\zeta=0)$.

Starting from $\zeta=0$, we can locally find holomorphic Darboux coordinates $\U^p(\zeta)$ and $\Ut_p(\zeta)$ ($p=1,...,n$, where $dim_\mathbb{R}M=4n$) for $\Omega_H(\zeta)$ that are analytic in $\zeta$ such that \cite{properties}
\beq \Omega_H(\zeta)=i\,d\U^p(\zeta)\wedge d\Ut_p(\zeta). \eeq{Darboux}
These canonical coordinates $\U,\Ut(\zeta)$ for $\Omega_H$ are crucial for the projective superspace formulation of $\sigma$-models with eight real supercharges\footnote{If we define $\breve{\U}(\zeta):=\Ub(-\frac{1}{\zeta})$, $\breve{\Ut}(\zeta):=\Utb(-\frac{1}{\zeta})$, then $(\U,\Ut)$ and $(\breve{\U}$,$\breve{\Ut})$ are related by a $\zeta^2$-twisted symplectomorphism whose generating function $f(\U,\breve{\U};\zeta)$ can be interpreted as the projective superspace Lagrangian \cite{properties}. \label{Footnote}}, where they describe "arctic" superfields. They have been determined for instance in \cite{properties} for the Eguchi-Hanson metric and we will use them in this paper to determine the generalized K\"ahler potential for hyperk\"ahler manifolds.

\section{Gen. K\"ahler structures on hyperk\"ahler manifolds}\label{sectionstructures}
We want to transport the idea of a twistor space from hyperk\"ahler to generalized K\"ahler geometry, namely we interpret a hyperk\"ahler manifold $(M,g,I,J,K)$ as a generalized K\"ahler manifold $(M,g,J_+,J_-)$, where we fix the left complex structure $J_+=I$ and let the right complex structure depend on $\zeta$: $J_-=\mathcal{J}(\zeta)$ (see eq. \eqref{J}). So for a given hyperk\"ahler manifold, we consider an $S^2$-family of generalized K\"ahler structures whose generalized K\"ahler potentials we now try to determine.

First, we need an explicit expression for the symplectic form $\Omega_G$ (eq. \eqref{omegag}), which now depends on $\zeta$. The anticommutator of two complex structures on a locally irreducible hyperk\"ahler manifold is equal to a constant times the identity, $\{J_+,J_-\}=c\mathbbm{1}$ (see, e.g., \cite{lectures}). If $J_+\neq\pm J_-$, then $|c|<2$ and $\frac{1}{\sqrt{4-c^2}}[J_+,J_-]$ is another complex structure, so in particular it squares to $-\mathbbm{1}$. Using this, we have
\beq \Omega_G=g[J_+,J_-]^{-1}=-\frac{1}{4-c^2}g[J_+,J_-], \eeq{OmGen}
which in our case, where we have $c=-2v_1=-2\frac{1-\zeta\zetab}{1+\zeta\zetab}$ and $[J_+,J_-]=2v_2K-2v_3J$, gives
\beq \Omega_G(\zeta)=-\frac{1}{2-2v_1^2}\left(v_2\omega_3-v_3\omega_2\right)=-\frac{1+\zeta\zetab}{8\zeta\zetab}\left[(\zeta+\zetab)\omega_3-i(\zetab-\zeta)\omega_2\right]. \eeq{OG}
This can be split into the sum of the holomorphic form $\Omega_+^{(2,0)}=i\zetab\frac{1+\zeta\zetab}{8\zeta\zetab}\omega^{(2,0)}$ and the antiholomorphic form $\Omega_+^{(0,2)}=-i\zeta\frac{1+\zeta\zetab}{8\zeta\zetab}\omega^{(0,2)}$ with respect to $J_+$ (see equation \eqref{omegapm}). Combining equations \eqref{OmegaH} and \eqref{Darboux}, we can choose the following Darboux coordinates for $\Omega_G(\zeta)$:
\ber x_L^p&=&\U^p(\zeta=0), \quad {y_L}_p=-\zetab\frac{1+\zeta\zetab}{8\zeta\zetab}\Ut_p(\zeta=0); \nonumber \\
\bar{x}_L^p&=&\Us^p(\zeta=0), \quad \bar{y}_{L_p}=-\zeta\frac{1+\zeta\zetab}{8\zeta\zetab}\Uts_p(\zeta=0). \eer{xyL}
With respect to $J_-=\mathcal{J}(\zeta)$, $\Omega_G$ splits into the sum of $\Omega_-^{(2,0)}=i\zetab\frac{1}{8\zeta\zetab}\Omega_H(\zeta)$ and $\Omega_-^{(0,2)}=-i\zeta\frac{1}{8\zeta\zetab}\overline{\Omega_H(\zeta)}$. Consequently, we can choose\footnote{We denote the complex conjugate of $\U(\zeta)$ by $\Us\equiv\Us(\zetab)\equiv\overline{\U(\zeta)}$ which is not to be confused with the notation in \cite{properties}, where $\Us$ is shorthand for $\breve{\U}(\zeta)=\Us(-\zeta^{-1})$.}
\ber x_R^p&=&\U^p(\zeta),\quad {y_R}_p=-\zetab\frac{1}{8\zeta\zetab}\Ut_p(\zeta); \nonumber \\
\bar{x}_R^p&=&\overline{\U^p(\zeta)},\quad {\yb}_{R_p}=-\zeta\frac{1}{8\zeta\zetab}\overline{\Ut_p(\zeta)}. \eer{xyR}
We are thus able to express the coordinates $x_{L,R}$ and $y_{L,R}$ that describe semichiral superfields in $\mathcal{N}=(2,2)$ models in terms of the coordinates $\U(\zeta)$, $\Ut(\zeta)$ describing arctic superfields in the projective superspace formulation of $\mathcal{N}=(4,4)$ supersymmetric sigma models. This will enable us to determine the $\zeta$-dependent generalized K\"ahler potential for hyperk\"ahler manifolds whose projective superspace description is known.

\section{The four-dimensional case \label{sectionfour}}

In this section, we consider the four-dimensional case and explicitly determine the partial differential equations for $\U(\zeta)$ and $\Ut(\zeta)$ in order to be holomorphic w.r.t. $\mathcal{J}(\zeta)$ and to fulfill equation \eqref{Darboux}. A four-dimensional K\"ahler manifold $(M,g,I)$ is hyperk\"ahler if and only if around each point there are holomorphic coordinates $(z,u)$ on $M$ such that the K\"ahler potential $K(z,u)$ fulfills the following Monge-Amp\`ere equation \cite{newman}:
\beq K_{u\ub}K_{z\zb}-K_{u\zb}K_{z\ub}=1. \eeq{Monge}
From a K\"ahler potential fulfilling this equation, we can construct the three K\"ahler forms:
\ber
\omega_1&=&-\frac{i}{2}\partial\bar{\partial} K, \nonumber \\
\omega_2&=&\frac{i}{2}(dz\wedge du-d\zb\wedge d\ub), \nonumber \\
\omega_3&=&\frac{1}{2}(dz\wedge du+d\zb\wedge d\ub).
\eer{kForms}
Together with the metric $g$ whose line element is
\beq ds^2=K_{u\ub}\,du d\ub+K_{u\zb}\,du d\zb+K_{z\ub}\,dz d\ub+K_{z\zb}\,dz d\zb, \eeq{metric}
we get the three complex structures, where equation \eqref{Monge} ensures that $J=g^{-1}\omega_2$ and $K=g^{-1}\omega_3$ indeed square to $-\mathbbm{1}$.

We find the following basis for the $(1,0)$ forms w.r.t. $\mathcal{J}(\zeta)$:
\beq \theta^1=dz-\zeta K_{u\ub} d\ub-\zeta K_{u\zb} d\zb,\quad \theta^2=du+\zeta K_{z\ub} d\ub+\zeta K_{z\zb} d\zb. \eeq{basisForms}
For $\U,\Ut$ to be holomorphic w.r.t. $\mathcal{J}(\zeta)$, $d\U(\zeta)$ and $d\Ut(\zeta)$ must be linear combinations of $\theta^1$ and $\theta^2$ (here the differential does not act on $\zeta$):
\beq d\U=\frac{\partial\U}{\partial z} \theta^1+\frac{\partial \U}{\partial u} \theta^2,\quad d\Ut=\frac{\partial\Ut}{\partial z} \theta^1+\frac{\partial \Ut}{\partial u}\theta^2. \eeq{sdgdfklj2}
Here the coefficients have been determined by comparing the $dz$- and $du$-terms on both sides. From equations \eqref{basisForms} and \eqref{sdgdfklj2}, we get the requirement that both $\U$ and $\Ut$ have to fulfill the following two PDEs:
\beq \frac{\partial \Psi}{\partial \zb}=\zeta\left(K_{z\zb} \frac{\partial \Psi}{\partial u}- K_{u\zb} \frac{\partial \Psi}{\partial z}\right),\quad \frac{\partial \Psi}{\partial \ub}=\zeta\left(K_{z\ub}\frac{\partial \Psi}{\partial u}-K_{u\ub} \frac{\partial \Psi}{\partial z}\right)\quad (\Psi=\U,\Ut).\eeq{requirement1}
Furthermore, we find that
\beq \Omega_H(\zeta)= idz\wedge du+i\zeta \partial \bar{\partial} K+i\zeta^2 d\zb\wedge d\ub=i\theta^1\wedge\theta^2, \eeq{afsdf}
so using \eqref{sdgdfklj2}, we obtain that equation \eqref{Darboux} corresponds to the requirement
\beq \frac{\partial \U}{\partial z}\frac{\partial \Ut}{\partial u}-\frac{\partial \U}{\partial u}\frac{\partial \Ut}{\partial z}=1.\eeq{requirement2}

\section{Euclidean space}\label{sectioneuclid}
We now use the relation between $x,y$ and $\U,\Ut$ derived in section \ref{sectionstructures} to determine the generalized K\"ahler potential for Euclidean space. Here the K\"ahler potential is given by
\beq K=u\ub+z\zb, \eeq{kEuclid}
which clearly fulfills equation \eqref{Monge}. Assuming that $(z,u)$ are holomorphic coordinates w.r.t. $I$ and setting $\omega^{(2,0)}=idz\wedge du$, we get the complex structures as described in section \ref{sectionfour}. They are the differentials (pushforwards) of the left action of the imaginary basis quaternions\footnote{We stick to the convention from previous papers and include the $i$-factor in the choice of $\omega^{(2,0)}$. This interchanges the complex structures $J$ and $K$, s.t. $J$ corresponds to left multiplication by $k$ and $K$ corresponds to left multiplication by $-j$.} $i,j,k$ on $\mathbb{H}\approx \mathbb{C}^2$, where we make the identification $(z,u)=(x_0+ix_1,x_2+ix_3)\mapsto x_0+ix_1+jx_2+kx_3$.
\beq \U(\zeta)=z-\zeta \ub,\quad \Ut(\zeta)=u+\zeta \zb \eeq{uFlat}
fulfill equations \eqref{requirement1} and \eqref{requirement2}, i.e. they are holomorphic w.r.t. $\mathcal{J}(\zeta)$ and satisfy equation \eqref{Darboux}.
Using equations \eqref{xyL} and \eqref{xyR}, we make the identifications
\beq x_L=\U(\zeta=0)=z,\,\,x_R=\U(\zeta)\equiv \U\text{ and }y_L=-\frac{1+\zeta\zetab}{8\zeta}u,\,\,y_R=-\frac{1}{8\zeta}\Ut. \eeq{identification4d}
Solving for $y_L$, $y_R$ in terms of $x_L$, $x_R$, we get
\beq y_L=-\frac{1+\zeta\zetab}{8\zeta\zetab}(\xb_L-\xb_R),\quad y_R=-\frac{1}{8\zeta\zetab}\left((1+\zeta\zetab)\xb_L-\xb_R\right),\eeq{sdfkjshgkjfdgc}
which leads (up to an additive constant) to the generating function (see equation \eqref{lrRelation})
\beq P=-\frac{1}{8\zeta\zetab}\left[x_R\xb_R+(1+\zeta\zetab)\cdot(x_L\xb_L-x_L\xb_R-\xb_Lx_R)\right]. \eeq{potflat}
This is the generalized K\"ahler potential for Euclidean space, where $J_+=I$ and $J_-$ is an arbitrary point on the twistor-sphere of complex structures, $J_-\neq \pm I$. However, we notice that $P$ only involves the combination $\zeta\zetab$, i.e. it only depends on the angle between $J_+$ and $J_-$ in the space spanned by the three complex structures $(I,J,K)$. Also $P$ turns out to be asymmetric between left- and right-coordinates. This can be resolved however, as there are various ambiguities in the generalized K\"ahler potential. For instance, we could distribute factors differently in \eqref{identification4d} or even perform a more complicated symplectomorphism, going to new coordinates $x'_{L/R}$, $y'_{L/R}$. If we make the identifications
\beq x'_L=i\sqrt{\frac{1+\zeta\zetab}{8\zeta}}z,\,\,x'_R=i\sqrt{\frac{1}{8\zeta}}\U\text{ and }y'_L=i\sqrt{\frac{1+\zeta\zetab}{8\zeta}}u,\,\,y'_R=i\sqrt{\frac{1}{8\zeta}}\Ut, \eeq{identification24d}
the potential is left-right-symmetric. Furthermore, we can perform Legendre transforms and express the potential in terms of a different set of variables. If we use for instance \eqref{identification24d} and in addition exchange the roles of $x'_R$ and $y'_R$, we arrive at the potential
\beq P'=\sqrt{1+\zeta\zetab}\cdot(x'_Ly'_R+\xb'_L\yb'_R)+\sqrt{\zeta\zetab}\cdot(x'_L\xb'_L+y'_R\yb'_R). \eeq{potentialprime}

\section{Example: Eguchi-Hanson geometry}\label{sectioneguchi}

The real function
\beq K=\sqrt{1+4u\ub(1+z\zb)^2}+\frac{1}{2}\text{log}\left[\frac{4u\ub(1+z\zb)^2}{\left(1+\sqrt{1+4u\ub(1+z\zb)^2}\right)^2}\right] \eeq{Kpot}
in the two complex variables $z,u$ fulfills the Monge-Amp\`ere equation \eqref{Monge}. It thus defines a hyperk\"ahler metric, where the K\"ahler forms are given by equation \eqref{kForms}. The first K\"ahler form takes the form
\ber \omega_1&=&-\frac{i}{2}\frac{1+z\zb}{\sqrt{1+4u\ub(1+z\zb)^2}}\Bigg[(1+z\zb)\,du\wedge d\ub+2u\zb \,dz\wedge d\ub	\nonumber \\
&&\quad\quad\quad\quad\quad+2z\ub \,du\wedge d\zb+\left(\frac{1}{(1+z\zb)^3}+4u\ub\right)dz\wedge d\zb\Bigg],
\eer{formHanson}
from which the metric can be read off. This is the well-known Eguchi-Hanson geometry\footnote{Setting $u=\frac{1}{2}u'^2$, $z=\frac{z'}{u'}$ and $r:=\sqrt{u'\ub'+z'\zb'}$ gives the familiar K\"ahler potential $K=\sqrt{1+r^4}+\text{log} \frac{r^2}{1+\sqrt{1+r^4}}$ for the Eguchi-Hanson metric  \cite{properties}.}. The holomorphic Darboux coordinates for $\Omega_H(\zeta)$ (fulfilling equations \eqref{requirement1} and \eqref{requirement2}) can be chosen as \cite{properties}
\ber \Ut&=&u+\zeta^2\zb^2\ub+\frac{\zb\zeta}{1+z\zb}\sqrt{1+4u\ub(1+z\zb)^2},\nonumber \\
\U&=&z-\frac{2\ub\zeta (1+z\zb)^2}{1+\sqrt{1+4u\ub(1+z\zb)^2}+2\ub\zb\zeta(1+z\zb)}. \eer{Upsilon}
We solve $\U,\Us(z,\zb,u,\ub)$ for $u$ and $\ub$ to get
\beq u(z,\zb,\U,\Us)=\frac{\zeta}{1+z\zb}\cdot\frac{(\zb-\Us)(1+\U\zb)}{\zeta\zetab(1+\Us z)(1+\U\zb)-(z-\U)(\zb-\Us)} \eeq{U}
and its complex conjugate. Using this and the identifications derived in section \ref{sectionstructures} (equation \eqref{identification4d}), we get $y_L(x_L,x_R)$. We then integrate $y_L(x_L,x_R)$ w.r.t. $x_L$ to get the generalized K\"ahler potential up to a possible additive term that is independent of $x_L$:
\ber P&=&\int\,y_L(x_L,x_R)\,dx_L=-\zetab\frac{1+\zeta\zetab}{8\zeta\zetab}\int\,u(z,\U)\,dz \nonumber \\
&=&-\frac{1}{8}\cdot \text{log}\frac{1+x_L\xb_L}{\zeta\zetab(1+x_L\xb_R )(1+\xb_Lx_R)-(x_L-x_R)(\xb_L-\xb_R)} \eer{potentialEH}
Plugging $u(z=x_L,\U=x_R)$ into $\Ut(z=x_L,u)$ (equation \eqref{Upsilon}) gives
\beq y_R(x_L,x_R)=-\zetab\frac{1}{8\zeta\zetab}\Ut(x_L,x_R)=-\frac{1}{8}\cdot\frac{(1+\zeta\zetab)\cdot \xb_L-\left(1-\zeta\zetab\cdot x_L\xb_L\right)\xb_R}{\zeta\zetab(1+x_L\xb_R )(1+\xb_Lx_R)-(x_L-x_R)(\xb_L-\xb_R)} \eeq{yr}
which is indeed equal to $-\frac{\partial P}{\partial x_R}$. $P$ is real, so $\frac{\partial P}{\partial \xb_L}=\yb_L$ and $\frac{\partial P}{\partial \xb_R}=-\yb_R$ are also fulfilled and thus equation \eqref{potentialEH} gives indeed the $\zeta$-dependent generalized K\"ahler potential for the Eguchi-Hanson geometry:
\beq 
P(x_L,\xb_L,x_R,\xb_R)=-\frac{1}{8}\cdot \text{log}\frac{1+|x_L|^2}{\zeta\zetab\cdot|1+x_L\xb_R |^2-|x_L-x_R|^2}. \eeq{potentialEH2}
Again, the generalized K\"ahler potential turns out to depend only on the combination $\zeta\zetab$, i.e. on the angle between $J_+$ and $J_-$. Of course, there are again many ambiguities in the potential, but \eqref{potentialEH2} seems to be already in its simplest form.

\section{Hyperk\"ahler structures on cotangent bundles of K\"ahler manifolds and projective superspace}\label{sectionproj}
The target space of 4D $\mathcal{N}=2$ sigma models is constrained to be a hyperk\"ahler manifold \cite{freedman}. This corresponds to $2D$ $\mathcal{N}=(4,4)$ sigma models without $B$-field\footnote{For vanishing $B$-field, all the results from $4D$ $\mathcal{N}=2$ projective superspace can be immediately transferred to $2D$ $\mathcal{N}=(4,4)$ projective superspace. Actually the results that we are using only depend on the target space geometry, not on the number of space-time dimensions.}. Projective superspace provides methods to construct such models and for a large class of examples it can be used to extract the arctic superfields $\U$ and $\Ut$ \cite{GatesKuzenko1},\cite{Arai1},\cite{Arai2} (see \cite{ulfProjective} for a review) that we need in order to determine the generalized K\"ahler potential of the hyperk\"ahler target space using the method derived in chapter \ref{sectionstructures}. This has been done in particular for the hyperk\"ahler structure on cotangent bundles of Hermitian symmetric spaces (\cite{GatesKuzenko1}-\cite{Arai3}). As an example, we use the results from \cite{GatesKuzenko1},\cite{Arai1},\cite{Arai2} to determine the generalized K\"ahler potential for $T^\ast\mathbb{C}P^n=T^\ast(SU(n+1)/U(n))$. The special case $n=1$ then corresponds to the Eguchi-Hanson geometry that we considered in the last section.

It is a well-known fact that a hyperk\"ahler metric exists on (some open subset of) the cotangent bundle of every K\"ahler manifold $M$ \cite{Feix},\cite{Kaledin},\cite{GatesKuzenko2}. In projective superspace, this corresponds to models where the projective superspace Lagrangian $f(\U,\breve{\U};\zeta)$ (see footnote \ref{Footnote}) does not explicitly depend on $\zeta$ \cite{GatesKuzenko1}. To obtain the coordinates $\U$ and $\Ut$ for $T^\ast M$, one takes the $\mathcal{N}=(4,4)$ projective superspace Lagrangian $f(\U,\breve{\U})$ to be the K\"ahler potential $K(\phi,\bar{\phi})$ of the base space $M$, where the arctic and antarctic superfields $\U,\breve{\U}$ replace the chiral and antichiral superfields $\phi,\bar{\phi}$ of the $\mathcal{N}=(2,2)$ description, i.e. the holomorphic coordinates on $M$. In $\mathcal{N}=(2,2)$ components, $\U$ decomposes as
\beq \U=\phi+\zeta \Sigma+\sum_{j=2}^\infty \zeta^j X_j,\eeq{sdfcvfwe}
where $\phi$ is a chiral superfield describing the coordinates on $M$, $\Sigma$ is a complex linear superfield describing coordinates in the fiber of the tangent bundle $TM$ and all higher order terms are unconstrained auxiliary superfields \cite{GatesKuzenko2},\cite{Arai3}. Solving the algebraic equations of motion for the auxiliary superfields yields $\U$, and thus the Lagrangian in terms of the $\mathcal{N}=(2,2)$ superfields $(\phi,\Sigma)$ and the auxiliary complex variable $\zeta$. Integrating out $\zeta$ and dualizing\footnote{The duality between chiral $\psi$ and complex linear superfields $\Sigma$ is just an ordinary coordinate transformation that does not change the target space geometry.} the action, i.e. performing a Legendre transform replacing the complex linear superfields $\Sigma$ by chiral superfields $\psi$, gives the transformation $\Sigma(\psi)$ from which we obtain $\U(\phi,\psi)$. Here, $\psi$ describes coordinates in the fiber of the cotangent bundle $T^\ast M$. $\Ut(\phi,\psi)$ can then be obtained from $f$ and $\U(\phi,\psi)$ via $\Ut=\zeta\frac{\partial f}{\partial \U}$ \cite{properties}. One can also read off the ordinary K\"ahler potential of $T^\ast M$ from the dualized action \cite{GatesKuzenko2},\cite{Arai3}.

\subsection{Generalized K\"ahler potential for $T^\ast\mathbb{C}P^n$}
The crucial step in the above procedure is to eleminate the infinite tower of unconstrained auxiliary $\mathcal{N}=(2,2)$ superfields using their algebraic equations of motion. This has been done for instance in \cite{Arai1} for $T^\ast\mathbb{C}P^n$.

For the projective sigma model with target space $T^\ast\mathbb{C}P^n$, we take the projective superspace Lagrangian $f$ to be the K\"ahler potential of the Fubini-Study metric on $\mathbb{C}P^n$:
\beq f(\U^i(\zeta),\breve{\U}^{i}(\zeta))=a^2 \text{log}\left(1+\frac{\U^j\breve{\U}^{j}}{a^2}\right). \eeq{kahlerpotentialcpn}
Here, $a$ is a real parameter. The equations of motion for the auxiliary superfields have been solved in \cite{Arai1}. This gives $\U$ in terms of chiral and complex linear $\mathcal{N}=(2,2)$ superfields\footnote{This result was already obtained in the preparation of \cite{GatesKuzenko2} and later independently derived and first published in \cite{Arai1}.}:
\beq \U^i=z^i+\zeta\frac{\Sigma^i}{1-\zeta\frac{\bar{z}^{\bar{k}}\Sigma^k}{a^2+ z^l\bar{z}^{\bar{l}}}}. \eeq{UInTermsOfPhiSigma}
Here, we change notation and let $z\equiv\phi$ parametrize the base space and $u\equiv\psi$ the fibers of the cotangent bundle.

Dualizing the action of the sigma model to go from complex linear coordinates $\Sigma$ to chiral coordinates $u$ gives the equations\footnote{Repeated indices are always summed over $1,...,n$ and the metric is always written out explicitly, i.e. we never use it to raise or lower indices.} \cite{Arai1}
\beq u_i=-\frac{g_{i\bar{j}}\bar{\Sigma}^{\bar{j}}}{1-\frac{g_{k\bar{l}}\Sigma^k\bar{\Sigma}^{\bar{l}}}{a^2}},\quad
\bar{u}_{\bar{i}}=-\frac{g_{\bar{i}j}\Sigma^j}{1-\frac{g_{k\bar{l}}\Sigma^k\bar{\Sigma}^{\bar{l}}}{a^2}};
\eeq{kjfdg}
which have to be solved for the old coordinates $\Sigma$, $\bar{\Sigma}$ in terms of $u$, $\bar{u}$. Here $g_{i\bar{j}}$ is
\beq g_{i\bar{j}}=\frac{a^2\delta_{ij}}{a^2+z^k\bar{z}^{\bar{k}}}-\frac{a^2\bar{z}^{\bar{i}}z^j}{(a^2+z^l\bar{z}^{\bar{l}})^2}, \eeq{FubiniStudyMetric}
the Fubini-Study metric on $\mathbb{C}P^n$, and $g^{i\bar{j}}$ is its inverse. We find the following solution:
\beq
\Sigma^i=-\frac{2g^{i\bar{j}}\bar{u}_{\bar{j}}}{1+\sqrt{1+4\frac{g^{k\bar{l}}u_k\bar{u}_{\bar{l}}}{a^2}}},\quad
\bar{\Sigma}^{\bar{i}}=-\frac{2g^{\bar{i}j}u_j}{1+\sqrt{1+4\frac{g^{k\bar{l}}u_k\bar{u}_{\bar{l}}}{a^2}}}.
\eeq{sdfsdg}
Plugging this into \eqref{UInTermsOfPhiSigma} gives the arctic superfields $\U$ in terms of chiral $\mathcal{N}=(2,2)$ superfields $z$ and $u$:
\beq \U^i=z^i-\zeta\frac{2\bar{u}_{\bar{j}}g^{i\bar{j}}}{1+\sqrt{1+4\frac{g^{k\bar{l}}u_k\bar{u}_{\bar{l}}}{a^2}}+2\zeta\frac{g^{p\bar{q}}\bar{z}^p\bar{u}_{\bar{q}}}{a^2+z^m\bar{z}^{\bar{m}}}}.\eeq{NeededUInTermsOfUZ}
Together with
\beq \Ut^i=\zeta\frac{\partial f}{\partial \U^i}=\frac{\zeta\breve{\U}^i}{1+\frac{\U^j\breve{\U}^{j}}{a^2}}, \eeq{UTilde}
this is all the information we need to determine the generalized K\"ahler potential for $T^\ast \mathbb{C}P^n$ using the identifications found in section \ref{sectionstructures}.

Solving \eqref{NeededUInTermsOfUZ} and its complex conjugate for $u$ and $\ub$ gives $u(z,\U)$:
\beq
u_i=\zeta \cdot \frac{a^2(a^2+z^{k}\zb^{\bar{k}})(a^2+\zb^{\bar{k}}\U^{k})\cdot g_{i\bar{j}}(\zb^{\bar{j}}-\Us^{\bar{j}})}{a^2\zeta\zetab(a^2+\zb^{\bar{k}} \U^{k})(a^2+z^{k}\Us^{\bar{k}})-(z^l-\U^l)g_{l\bar{m}}(\zb^{\bar{m}}-\Us^{\bar{m}})(a^2+z^{k}\zb^{\bar{k}})^2}.
\eeq{uOfZU}
Integrating this with respect to $z^i$ gives:
\beq \int u_i\,dz^i= \frac{\zeta a^2}{1+\zeta\zetab}\text{log}\frac{a^2+\mathbf{z}^T\mathbf{\zb}}{a^2\zeta\zetab(a^2+\mathbf{\zb}^T\mathbf{\U})(a^2+\mathbf{z}^T\mathbf{\Ub})-(\mathbf{z}-\mathbf{\U})^T\mathbf{g}(\mathbf{\zb}-\mathbf{\Us})(a^2+\mathbf{z}^T\mathbf{\zb})^2}. \eeq{sdgfdh}
Here, no sum is implied on the left-hand side and on the right-hand side we use vector notation ($\mathbf{z}:=(z^1,...,z^n)^T$, etc.) and $\mathbf{g}:=(g_{i\bar{j}})_{1\leq i,j\leq n}$. So, up to an additive term $c(x_R,\xb_R)$, the generalized K\"ahler potential for $T^*\mathbb{C}P^n$ is
\beq P=-\frac{a^2}{8}\text{log}\frac{a^2+\mathbf{x_L}^T\mathbf{\xb_L}}{a^2\zeta\zetab(a^2+\mathbf{\xb_L}^T\mathbf{x_R})(a^2+\mathbf{x_L}^T\mathbf{\xb_R})-(\mathbf{x_L}-\mathbf{x_R})^T\mathbf{g}(\mathbf{\xb_L}-\mathbf{\xb_R})(a^2+\mathbf{x_L}^T\mathbf{\xb_L})^2}. \eeq{finalpotential}
For $n=1$ and $a=1$, we have $g_{z\bar{z}}=\frac{1}{(1+z\bar{z})^2}$ and all the results from section \ref{sectioneguchi} are reproduced. Therefore, we assume that $c(x_R,\xb_R)$ can be set to zero.


\section{Discussion}
There is an increasing number of examples, most notably among Hermitian symmetric spaces, where the decomposition of the $\mathcal{N}=(4,4)$ arctic superfields $\U$, $\Ut$ in terms of their $\mathcal{N}=(2,2)$ components ($z,u$) has been determined. In these cases, one can apply the methods developed in this paper to determine more examples of generalized K\"ahler potentials on hyperk\"ahler manifolds. Having whole classes of manifolds available for our analysis, one could try to find more general statements about the generalized K\"ahler potential in the case of hyperk\"ahler manifolds. 

The Eguchi-Hanson geometry is one of the hyperk\"ahler manifolds that can be obtained from the generalized Legendre transform construction in \cite{hitchin} (generalized T-duality). The manifolds stemming from that construction are $4n$-dimensional hyperk\"ahler manifolds admitting $n$ commuting tri-holomorphic killing vectors. They are called toric hyperk\"ahler manifolds and have been classified in \cite{Bielawski}. It should be possible to determine the relevant coordinates $\U(z,u;\zeta)$ and $\Ut(z,u;\zeta)$ for toric hyperk\"ahler manifolds. For four-dimensional toric hyperk\"ahler manifolds, \cite{sevrin} gives a formula for the generalized K\"ahler potential as a certain threefold Legendre transform in the special case $\zeta\zetab=1$. One could compare this construction with our results at least for the examples given in this paper or try to relate the two methods in general for four-dimensional toric hyperk\"ahler manifolds. As a further explicit example, one could for instance consider the Taub-NUT geometry and determine its generalized K\"ahler potential.

The generalized K\"ahler potential for the Eguchi-Hanson geometry can also be obtained from a generalized quotient of Euclidean $8$-dimensional space by a $U(1)$-isometry and in this setting turns out to be exactly \eqref{potentialEH2} as well \cite{marcos}.

$2D$ $\mathcal{N}=(2,2)$ sigma models have a target space that is a generalized K\"ahler manifold and in general, they are described by chiral, twisted chiral and semichiral superfields. The models with hyperk\"ahler target space and $J_+\neq \pm J_-$ are described purely in terms of semichiral superfields. These models do not in general admit off-shell $\mathcal{N}=(4,4)$ supersymmetry, since it was shown in \cite{Malin1} and \cite{Malin2} that a sigma model parametrized by semichiral fields can only be extended to off-shell $\mathcal{N}=(4,4)$ supersymmetry if the target space is $4n$-dimensional with $n>1$. However, they are always dual to models with $\mathcal{N}=(4,4)$ supersymmetry that are parametrized by chiral and twisted chiral superfields \cite{sevrin}. The exact relation between the $\mathcal{N}=(4,4)$ sigma models described by chiral and twisted chiral superfields, and their dual semichiral models will be described in \cite{Malin3}.

The relation between the coordinates $x_{L/R}$, $y_{L/R}$ and $\U$, $\Ut$ has been obtained in this paper from a purely differential geometric approach. $x_{L/R}$, $y_{L/R}$ describe left- and right-semichiral superfields in $2D$ $\mathcal{N}=(2,2)$ sigma models. For a target space that is hyperk\"ahler, these models are dual to models with $\mathcal{N}=(4,4)$ supersymmetry that are parametrized by chiral and twisted chiral superfields. The coordinates $\U(\zeta)$, $\Ut(\zeta)$ however describe arctic superfields in $\mathcal{N}=(4,4)$ sigma models in projective superspace. The field theoretical interpretation and understanding of this relation between arctic $\mathcal{N}=(4,4)$ models and the semichiral models that are dual to $\mathcal{N}=(4,4)$ models remains an open problem. The complex coordinate $\zeta$ is an auxiliary variable that gets integrated out in projective superspace, but for ordinary superspace it is just a constant parameter. Thus in our relation, an arctic model corresponds to a two-sphere (or more precisely to a cylinder) of presumably equivalent semichiral models.

In this paper, we mainly focused on the special case, where the bihermitian structure $(J_+,J_-)$ only depends on one complex parameter $\zeta$ and established a relation to projective superspace. In the appendix, we show that the results from section \ref{sectionstructures} can be generalized to the full $S^2\times S^2$-family of generalized complex structures on a hyperk\"ahler manifold parametrized by two complex parameters $\zeta_+$ and $\zeta_-$. Many hints point towards an intimate relation of this formulation to doubly-projective superspace \cite{doublyProjective},\cite{doublyProjective2}.

\bigskip\bigskip
\noindent{\bf\Large Acknowledgements}:
\bigskip

\noindent
The author owes a lot to Martin Ro\v cek, who initiated the project and kept it alive, providing motivation and knowledge. The author would like to thank P. Marcos Crichigno for valuable discussions during the creation of this paper. Ulf Lindstr\"om and Malin G\"oteman have provided comments helping to put the paper into its final form.

\appendix
 \Section{Appendix: $S^2\times S^2$-family of generalized complex structures}
\setcounter{equation}{0}
In this appendix, we generalize our results from section \ref{sectionstructures}. Instead of fixing one complex structure, we can also let both $J_+$ and $J_-$ depend on an individual complex coordinate and thus consider an $S^2\times S^2$-family of generalized complex structures $(M,g,J_+,J_-)$ on a given hyperk\"ahler manifold $(M,g,I,J,K)$. We parametrize vectors $\vec{u}$, $\vec{v}\in S^2\backslash (-1,0,0)$ by complex coordinates $\zeta_+$, $\zeta_-$ like in equation \eqref{J} and define
\beq J_+:=\mathcal{J}(\zeta_+)=u_1I+u_2J+u_3K,\quad J_-:=\mathcal{J}(\zeta_-)=v_1I+v_2J+v_3K. \eeq{jpm}
The anticommutator depends only on the angle $\theta$ between $\vec{u}$ and $\vec{v}$:
\beq \{J_+,J_-\}=-2(\vec{u}\cdot\vec{v})\mathbbm{1}=-2\,\cos{\theta}\,\mathbbm{1}. \eeq{anticomm}
The commutator turns out to be perpendicular to $J_+$ and $J_-$ in the space spanned by $(I,J,K)$:
\beq [J_+,J_-]=2(u_2v_3-u_3v_2)I-2(u_1v_3-u_3v_1)J+2(u_1v_2-u_2v_1)K=2(\vec{u}\times\vec{v})\cdot (I,J,K)^T. \eeq{comm}

In order to determine the coordinates $x_R,y_R$, we need to split $\Omega_G$ into a $(2,0)$- and a $(0,2)$-form w.r.t. $J_-$. Indeed, we find that
\beq g[J_+,J_-]=\frac{i}{(1+\zeta_-\zetab_-)^2}\left((\overline{\vec{a}(\zeta_-)}\cdot \vec{u})\,\Omega_H(\zeta_-)-(\vec{a}(\zeta_-)\cdot\vec{u})\,\overline{\Omega_H(\zeta_-)}\right), \eeq{split}
where $\vec{a}(\zeta)=(-2\zeta,1-\zeta^2,i(1+\zeta^2))^T$, i.e. $\Omega_H(\zeta)=\vec{a}(\zeta)\cdot\vec{\omega}$ (see eq. \eqref{OmegaH}).
So we find
\beq \Omega_G=-\frac{1}{4-4(\vec{u}\cdot\vec{v})^2}g[J_+,J_-]=\Omega_-^{(2,0)}+\Omega_-^{(0,2)}, \eeq{dsfhgdh}
where
\beq
\Omega_-^{(2,0)}=\frac{-i\,(\overline{\vec{a}(\zeta_-)}\cdot \vec{u})}{4\sin^2{\theta}\,(1+\zeta_-\zetab_-)^2}\,\Omega_H(\zeta_-)\equiv-ic_-\Omega_H(\zeta_-),\quad \Omega_-^{(0,2)}=\overline{\Omega_-^{(2,0)}}.
\eeq{splitting2}
Thus knowing that $\Omega_H(\zeta_-)=id\U^p(\zeta_-)\wedge d\Ut_p(\zeta_-)$, we can choose (omitting indices)
\beq
x_R=\U(\zeta_-),\quad y_R=c_-\Ut(\zeta_-).
\eeq{ident2}
to get $\Omega_G=dx_R\wedge dy_R+d\xb_R\wedge d\yb_R$.

Exchanging the roles of $\vec{u}$,$\vec{v}$ and $\zeta_+$, $\zeta_-$ respectively and considering the antisymmetry of $[J_+,J_-]$, we get the following splitting w.r.t. $J_+$:
\beq
\Omega_+^{(2,0)}=\frac{i\,(\overline{\vec{a}(\zeta_+)}\cdot \vec{v})}{4\sin^2{\theta}\,(1+\zeta_+\zetab_+)^2}\,\Omega_H(\zeta_+)\equiv -ic_+\Omega_H(\zeta_+),\quad \Omega_+^{(0,2)}=\overline{\Omega_+^{(2,0)}},
\eeq{splitting3}
which allows us to choose
\beq
x_L=\U(\zeta_+),\quad y_L=c_+\Ut(\zeta_+).
\eeq{ident3}
The constants $c_+,c_-(\zeta_+,\zeta_-)$ can be written as
\beq c_+=\frac{1+\zeta_-\zetab_-}{8(1+\zeta_+\zetab_-)(\zeta_+-\zeta_-)}, \quad c_-=\frac{1+\zeta_+\zetab_+}{8(1+\zetab_+\zeta_-)(\zeta_+-\zeta_-)}.\eeq{cConstants}

We see that by exchanging $\zeta_+$ with $\zeta_-$, we exchange $x_L$ with $x_R$ and $y_L$ with $-y_R$.
In the special case $\zeta_+=0$ (i.e. $\vec{u}=(1,0,0)$) and $\zeta_-=\zeta$, we have $\sin^2\theta=\frac{4\zeta\zetab}{(1+\zeta\zetab)^2}$ and \eqref{ident2}, \eqref{ident3} reduce to the results \eqref{xyL}, \eqref{xyR} from section \ref{sectionstructures}.

Using \eqref{ident2} and \eqref{ident3}, we can now also determine an $S^2\times S^2$-family of generalized K\"ahler potentials $P_{\zeta_+,\zeta_-}$ for hyperk\"ahler manifolds. For Euclidean space, we find the $\zeta_+$- and $\zeta_-$-dependent generalized potential to be
\ber P&=&-\frac{1}{8(\zeta_+-\zeta_-)(\zetab_+-\zetab_-)}\Bigg[(1+\zeta_-\zetab_-)x_L\xb_L+(1+\zeta_+\zetab_+)x_R\xb_R \nonumber \\ &\,&\,\quad\quad\quad\quad\quad\quad\quad-(1+\zeta_+\zetab_+)(1+\zeta_-\zetab_-)\left(\frac{x_L\xb_R}{1+\zeta_+\zetab_-}+\frac{\xb_Lx_R}{1+\zetab_+\zeta_-}\right)\Bigg]. \eer{skhgdf3456}


\begin{thebibliography}{6666}

\bibitem{freedman}
  L.~Alvarez-Gaume and D.~Z.~Freedman,
  ``Geometrical Structure And Ultraviolet Finiteness In The Supersymmetric
  Sigma Model,''
  Commun.\ Math.\ Phys.\  {\bf 80}, 443 (1981).
  
\bibitem{hitchin}
  N.~J.~Hitchin, A.~Karlhede, U.~Lindstrom and M.~Rocek,
  ``Hyperkahler Metrics and Supersymmetry,''
  Commun.\ Math.\ Phys.\  {\bf 108}, 535 (1987).
  
\bibitem{properties}
  U.~Lindstrom and M.~Rocek,
  ``Properties of hyperkahler manifolds and their twistor spaces,''
  Commun.\ Math.\ Phys.\ \ {\bf 293}, 257  (2010)
  [arXiv:0807.1366 [hep-th]].
  
\bibitem{gualtieri}
M.~Gualtieri, ``Generalized complex geometry,'' Oxford University  
DPhil thesis,
[arXiv:math/0401221v1].

\bibitem{gates}
  S.~J.~.~Gates, C.~M.~Hull and M.~Rocek,
  ``Twisted Multiplets And New Supersymmetric Nonlinear Sigma Models,''
  Nucl.\ Phys.\  B {\bf 248}, 157 (1984).

\bibitem{offshell}
  U.~Lindstrom, M.~Rocek, R.~von Unge and M.~Zabzine,
  ``Generalized Kahler manifolds and off-shell supersymmetry,''
  Commun.\ Math.\ Phys.\ \ {\bf 269}, 833  (2007)
  [hep-th/0512164].
  
\bibitem{linearizing}
   U.~Lindstrom, M.~Rocek, R.~von Unge and M.~Zabzine,
  ``Linearizing Generalized Kahler Geometry,''
  JHEP\ {\bf 0704}, 061  (2007)
  [hep-th/0702126].
  
\bibitem{hitchinP}
N.~Hitchin,
  Commun.\ Math.\ Phys.\ \ {\bf 265}, 131  (2006)
  [math/0503432 [math-dg]].
  
\bibitem{lectures}
  A.~Moroianu,
  ``Lectures on K\"ahler geometry,''
  London Mathematical Society Student Texts {\bf 69},
  Cambridge University Press, 2007.

\bibitem{newman}
  S.~Chakravarty, L.~J.~Mason and E.~T.~Newman,
  ``Canonical structures on antiselfdual four manifolds and the diffeomorphism group,''
  J.\ Math.\ Phys.\ \ {\bf 32}, 1458  (1991).
  
\bibitem{sevrin}
  J.~Bogaerts, A.~Sevrin, S.~van der Loo and S.~Van Gils,
  ``Properties of semichiral superfields,''
  Nucl.\ Phys.\ B\ {\bf 562}, 277  (1999)
  [hep-th/9905141].
  
\bibitem{marcos}
  P.~M.~Crichigno,
  private communication, in preparation.
  
\bibitem{potential}
    U.~Lindstrom, M.~Rocek, R.~von Unge and M.~Zabzine,
  ``A potential for Generalized Kahler Geometry,''
  IRMA Lect.\ Math.\ Theor.\ Phys.\ \ {\bf 16}, 263  (2010)
  [hep-th/0703111].
  
\bibitem{Bielawski}
  R.~Bielawski,
  ``Complete hyperKahler 4n manifolds with n commuting triHamiltonian vector fields,''
  math/9808134.
    
\bibitem{Feix}
	B.~Feix,
	``Hyperk\"ahler metrics on cotangent bundles,''
	J. Reine Angew. Math. {\bf 532} (2001), 33-46
	
\bibitem{Kaledin}
	D.~Kaledin,
	``Hyperk\"ahler structures on total spaces of holomorphic cotangent bundles''
	[arXiv:alg-geom/9710026v1]\\
	D.~Kaledin,
	``A canonical hyperk\"ahler metric on the total space of a cotangent bundle''
	[arXiv:math/0011256v1]    

\bibitem{ulfProjective}
  U.~Lindstr\"om,
  ``Hyperkahler metrics from projective superspace,''
  arXiv:hep-th/0703181.
  
\bibitem{GatesKuzenko1}
  S.~J.~J.~Gates and S.~M.~Kuzenko,
  ``The CNM hypermultiplet nexus,''
  Nucl.\ Phys.\  B {\bf 543}, 122 (1999)
  [arXiv:hep-th/9810137].
  
\bibitem{GatesKuzenko2}
  S.~J.~J.~Gates and S.~M.~Kuzenko,
  ``4-D, N=2 supersymmetric off-shell sigma models on the cotangent bundles of
  Kahler manifolds,''
  Fortsch.\ Phys.\  {\bf 48}, 115 (2000)
  [arXiv:hep-th/9903013].
  
\bibitem{Arai1}
  M.~Arai and M.~Nitta,
  ``Hyper-Kahler sigma models on (co)tangent bundles with SO(n) isometry,''
  Nucl.\ Phys.\  B {\bf 745}, 208 (2006)
  [arXiv:hep-th/0602277].
  
\bibitem{Arai2}
  M.~Arai, S.~M.~Kuzenko and U.~Lindstr\"om,
  ``Hyperkahler sigma models on cotangent bundles of Hermitian symmetric spaces
  using projective superspace,''
  JHEP {\bf 0702}, 100 (2007)
  [arXiv:hep-th/0612174].

\bibitem{Arai3}
  M.~Arai, S.~M.~Kuzenko and U.~Lindstr\"om,
  ``Polar supermultiplets, Hermitian symmetric spaces and hyperkahler
  metrics,''
  JHEP {\bf 0712}, 008 (2007)
  [arXiv:0709.2633 [hep-th]].
    
\bibitem{Malin1}
  M.~Goteman and U.~Lindstrom,
  ``Pseudo-hyperkahler Geometry and Generalized Kahler Geometry,''
  Lett.\ Math.\ Phys.\ \ {\bf 95}, 211  (2011)
  [arXiv:0903.2376 [hep-th]].
  
\bibitem{Malin2}
  M.~Goteman, U.~Lindstrom, M.~Rocek and I.~Ryb,
  ``Sigma models with off-shell N=(4,4) supersymmetry and noncommuting complex structures,''
  JHEP\ {\bf 1009}, 055  (2010)
  [arXiv:0912.4724 [hep-th]].
  
\bibitem{Malin3}
M.~G\"oteman,
  in preparation.

\bibitem{doublyProjective}
  T.~Buscher, U.~Lindstrom and M.~Rocek,
  ``NEW SUPERSYMMETRIC sigma MODELS WITH WESS-ZUMINO TERMS,''
  Phys.\ Lett.\ B\ {\bf 202}, 94  (1988).

\bibitem{doublyProjective2}
  U.~Lindstrom, I.~T.~Ivanov and M.~Rocek,
  ``New N=4 superfields and sigma models,''
  Phys.\ Lett.\ B\ {\bf 328}, 49  (1994)
  [hep-th/9401091].
\end{thebibliography}
\end{document}